\documentclass[preprint,12pt,onecolumn,byrevtex,aps,showpacs]{revtex4}%
\usepackage{amsfonts}
\usepackage{amsmath}
\usepackage{amssymb}
\usepackage{graphicx}%
\begin{document}
\title{Non-Gaussian conductance noise in disordered electronic systems due to a
non-linear mechanism}
\author{V. Orlyanchik$^{1}$, V. I. Kozub$^{1,2}$, and Z. Ovadyahu$^{1}$\vspace{0.2in}}
\affiliation{$^{1}$Racah Institute of Physics, The Hebrew University, Jerusalem, 91904,
Israel, $^{2}$A. F. Ioffe Physico Technical Institute, St. Petersburg 194021,
Russia\vspace{0.01in}.}

\begin{abstract}
We present results of conductance-noise experiments on disordered films of
crystalline indium oxide with lateral dimensions $2\mu m$ to $1mm.$ The
power-spectrum of the noise has the usual $1/f$ $~$form, and its magnitude
increases with inverse sample-volume down to sample size of $2\mu m,$ a
behavior consistent with un-correlated fluctuators. A colored second spectrum
is only occasionally encountered (in samples smaller than $40\mu m$), and the
lack of systematic dependence of non-Gaussianity on sample parameters
persisted down to the smallest samples studied ($2\mu m$). Moreover, it turns
out that the degree of non-Gaussianity exhibits a non-trivial dependence on
the bias $V~$used in the measurements$;$ it initially increases with $V$ then,
when the bias is deeper into the non-linear transport regime it decreases with
$V$. We describe a model that reproduces the main observed features and argue
that such a behavior arises from a non-linear effect inherent to electronic
transport in a hopping system and should be observed whether or not the system
is glassy.

\end{abstract}
\maketitle

\section{Introduction}

Conduction noise is a property of essentially all electronic systems. A common
form of this noise has a $1/f^{\eta}$ power-spectrum with $\eta$ of order
unity \cite{1}. Such a spectrum may result from the superposition of many
fluctuators with individual frequencies $\omega_{i}$ extending uniformly over
the observed range \cite{1}. When these fluctuators are independent, the
central limit theorem mandates that the power-spectrum be Gaussian, and this
is presumably the generic result in the thermodynamic limit. Therefore when a
system with power-spectrum of the $1/f$ ~type and a non-Gaussian nature is
encountered, it seems reasonable to conclude that the associated fluctuators
are correlated.

Identifying the underlying source of the correlations however, is not a
trivial task as there could be several reasons for non-Gaussian noise,
including artifacts, and correlations in the noise may appear even when the
fluctuators are basically independent. For example, Seidler \textit{et al}
\cite{2}\textit{,} using their dynamical-current-redistribution (DCR) model,
argued that a non-Gaussian noise may result from features that are peculiar to
the process of electrical conductivity in a network; Their considerations are
especially relevant for inhomogeneous systems, and when the fluctuations are
strong. Both ingredients are inherent to transport in disordered conductors
where non-Gaussianity were frequently observed \cite{3}. The archtypical
system of this class is a hopping system where transport is confined to a
percolation network. Such a system may be viewed as a random-resistor-network
in which each resistor $r_{ij}~$is associated with pair of sites $i,j$ that
are connected by a hopping process \cite{4}. The wide distribution of the
$r_{ij}$'s in the random-resistor-network leads to several unique features of
electronic transport in such a medium. The most familiar feature is that the
current in the system is carried by a percolation-network \cite{5,6,7}
involving a relatively small number of `critical' resistors. This percolation
problem differs from the classical, `geometric' scenario in two essential
accounts; First, the current-carrying network (CCN) of the hopping system is
temperature dependent - it becomes progressively more rarefied as temperature
decreases. Secondly, the resistances that comprise the CCN, as well as
elemental fluctuators that modulate them, typically exhibit non-linear
effects. At low temperatures, these non-linear effects are quite prominent
even for small applied fields. As we shall see, these features introduce a new
set of considerations into the question of noise correlations.

In this paper we report on a study of noise statistics of crystalline
indium-oxide films ($In_{2}O_{3-x}$) with their disorder tuned to make them
strongly localized. Previous studies of $1/f$ noise using $In_{2}O_{3-x}$
films has focused on the \textit{metallic} regime and employed macroscopic
samples \cite{8}. The emphasis in the current study is on the results obtained
with samples that exhibit prominent mesoscopic effects. The other aspect that
distinguishes this work from the previous study (where room temperature
measurements were employed) is that the noise experiments were carried out at
liquid helium temperatures. The original motivation for these experiments was
an attempt to find the spatial extent of the correlation length in the glassy
phase of $In_{2}O_{3-x}$ films by monitoring the noise characteristics of such
films, in particular, the second-spectrum \cite{9}, as function of their size.
The rationale being that below a certain sample size the correlated nature of
the electron glass should manifest itself in a respective correlation of the
conductance fluctuations. That should presumably occur once the spatial scale
associated with the glassy effects exceed the sample size. Samples with
lateral dimensions ranging from 1mm down to 2$\mu m$ were studied for this
purpose. A colored second spectrum, indicative of correlations, was indeed
found for samples with sizes at the lower part of this range. However, the
phenomenology encountered in these measurements was not in line with the
expectation based on the above scenario. In particular, even in the small size
regime, the occurrence of colored second spectrum was not consistent; samples
with nearly identical parameters gave conflicting results. This led us to
suspect that the source of correlations is not related to the underlying
glass. Further experiments revealed that that the degree of coloration in the
second spectrum depends in a non trivial way on the bias $V$ used in the
measurements. We propose a heuristic picture that qualitatively accounts for
these observations. This model is generic to transport in disordered systems
where transport is inherently inhomogeneous, and does not depend on the system
being glassy.

\section{Samples preparation and measurements techniques}

The $In_{2}O_{3-x}$ films used here were e-gun evaporated on either a $110\mu
m$ thick microscope-cover glass, or on a $\operatorname{Si}O_{2}$ insulating
layer ($0.5\mu m$ thick) thermally grown on a Si wafer. The latter was boron
doped and had resistivity $2\cdot10^{-3}\Omega cm,$ deep in the degenerate
regime. It thus could be used as the gate electrode for a low-temperature
measurement with the sample configured as a MOSFET device in which a thin film
of $In_{2}O_{3-x}$ served as the active layer. Lateral dimensions of the
samples were controlled using a stainless steel mask (for samples larger than
$0.5mm$), or optical lithography (for sizes in the range $30-200\mu m$) and
e-beam lithography for samples smaller than $30\mu m$. Samples used in this
study had length ($L$) and width ($W$) that ranged from $2mm$ down to $2\mu m$
and typical thickness $d=55\pm5\mathring{A}$. The `source' and `drain'
contacts were made from thermally evaporated $\approx500\mathring{A}$ thick
gold films. Fuller details of sample preparation and characterization are
given elsewhere \cite{10}.

Conductance measurements were performed using two terminal ac technique,
employing ITHACO-1211 current preamplifier and PAR-124A lock-in amplifier. In
the MOSFET-like samples, gate voltage sweeps were affected by charging a
$10\mu F$ capacitor with a constant current source (Keithley 220). All the
measurements reported here were performed with the samples immersed in liquid
helium at $T=4.11$K held by a 100 liters storage-dewar. This allowed long term
measurements without disturbing the samples as well as a convenient way to
maintain a stable bath temperature. These requirements are of particular
importance for studies of glassy systems where sample history may influence
time dependent measurements, as was demonstrated in previous studies using
such samples \cite{10}. To cater for this, samples were allowed to equilibrate
for at least 15 hours prior to any conductance versus time measurement. No
change in the nature of the noise of a given sample was found when
measurements were repeated a week after the initial cool-down.\texttt{ }

A typical hopping length at these temperatures, and for the range of
resistances used in this work is $\approx200\mathring{A}$ \cite{11}, which
makes such samples effectively two-dimensional (2D). This fact will be used in
the proposed model described in section IIIC.

Noise measurements employed a two terminal technique. These were performed by
biasing the sample with a constant (dc) voltage source (home-made rechargeable
battery-stack) while measuring the resulting current fluctuations by the ac
voltage drop $V_{ac}$ detected across a series resistor. This $V_{ac}$ was
monitored by a Dynamic Signal Analyzer (HP35670A) buffered by EG\&G5113
low-noise pre-amplifier. These data were then used to calculate the first and
second power-spectrum. The latter were implemented by the method suggested by
Restle \textit{et al} \cite{12}. As will be demonstrated below, the noise
magnitude of the samples studied here is quite large. This is due to three
reasons; (1) The samples reported here are strongly disordered and exhibit
large noise even when of macroscopic size \cite{8}. (2) The samples are
physically small, c.f., figure 2. (3) At the measured temperature, the samples
are rather deep into the hopping regime, which means that the effective volume
for the current carrying process in these samples may be considerably smaller
than their geometrical volume. These factors combine to give a large noise
magnitude, which made it unnecessary to use a more comprehensive measurement
configuration such as a 5-probe technique (that is awkward to employ for
insulating samples being inherently inhomogeneous). We did however check that
the contacts do not contribute to the noise by measuring few samples using a
4-probe technique.

The stability of the temperature bath afforded by the 100 liter storage dewar
was sufficiently accurate to neglect time-dependent temperature effects on the
samples conductance relative to the inherent 1/f noise. This was ascertained
by monitoring the resistance fluctuations of a Ge thermometer attached to the
sample stage as in Fig.5 of Vaknin et al \cite{10}.

\section{Results and discussion}

\subsection{Preliminary measurements}

A conductance time-trace and associated power-spectrum for the smallest sample
in the series studied are shown in Fig.1. Apart from its large magnitude
(presumably due to the small sample size), the noise has similar
characteristics as these of metallic samples of this material \cite{8}, in
particular the overall shape of the power-spectrum retains its $1/f^{\eta}$
character although some deviations from a power-law may be observed; a
best-fit to the data yields $\eta\simeq1.1$ but the low $f$ region seems to
follow a faster dependence. This is not surprising as such samples contain a
rather small number of fluctuating resistances as will be shown later. The
dependence of the noise magnitude on sample volume is shown in Fig.2 for
several of the studied samples. To facilitate comparison, the figure includes
only samples that are all of the same thickness, and have similar resistances.
Note that the noise magnitude is inversely proportional to volume down to
sample size of $2x2\mu m.$ This result is consistent with what is expected of
ensemble-averaged \textit{un}-correlated fluctuators. It might therefore
suggest that, if there are correlations in the noise in this system, they
occur on scales smaller than $2\mu m.$ Surprisingly then, when analyzing
preliminary data for samples in this series we encountered several traces that
yielded a frequency-dependent second-spectrum. Such occurrences were not
encountered in samples with sizes $\geq100\mu m,$ and they were still rare in
the range of sizes $30-50\mu m$. Colored second-spectrum could be observed in
samples as large as $\approx40\mu m$ but in unpredictable way; Two samples
with nearly identical parameters (resistance, size), and measured under
similar conditions, gave conflicting results; one exhibited
frequency-dependent second spectrum, while the other had a Gaussian spectrum.
An example of such a `conflicting couple' is shown in Fig.3 for $30x40\mu m$
samples. This lack of systematic behavior persisted down to our smallest
sample sizes, although the frequency of a colored second-spectrum appearances
seem to grow with decreasing size.

In contrast with the inconsistent appearance of noise correlations, all
samples in this series systematically show the same glassy features as
macroscopic samples with similar parameters. For example, Fig.4 shows the
`memory' cusp, which is the earmark of the electron-glass \cite{10}, for the
$30x40\mu m$ sample shown in Fig.3. Note that this actually is the sample that
exhibits \textit{Gaussian} noise. It is therefore unlikely that the
$f$-dependent second spectrum can be related to the correlations due to the
interactions that are associated with electron-glass dynamics. This conclusion
will be re-enforced by a more elaborate analysis and the ensuing discussion below.

In addition to the glassy cusp, the data in Fig.4 show mesoscopic conductance
fluctuations (CF). These are the `fingerprints' of the underlying CCN, and
reflect the process by which some critical resistors in the CCN are replaced
by other critical resistors as the chemical potential is varied \cite{13}. The
relative magnitude of the CF is a function of $\mathcal{L}$, the correlation
length of the percolation network. This is based on two assumptions: First,
the basic conductance swing $\Delta G$ associated with a critical resistor is
of the order of its conductance $G.$ This is a characteristic feature of the
strongly localized regime, (which for the present case of 2D samples means
sheet resistance $R_{\square}$ that fulfills~$R_{\square}\gg\frac{\hbar}%
{e^{2}}$) \cite{14}. Secondly, the relative fluctuation amplitude $\Delta G/G$
for the entire sample is essentially determined by the square root of the
number of critical resistors in the sample $N\approx\frac{LW}{\mathcal{L}^{2}%
},$ therefore, $\frac{\Delta G}{G}\approx\sqrt{\frac{\mathcal{L}^{2}}{LW}}$.
This relation will be used in this paper to estimate $\mathcal{L}~$as well as
it dependence on applied voltage$.$

\subsection{Dependence on applied field}

Low temperature transport measurements on hopping systems are very sensitive
to the value of the voltage $V$ used. It is notoriously difficult to achieve
linear response conditions in such cases, especially for small samples at low
temperatures, and deviations from Ohm's law are hard to avoid. This is
illustrated in Fig.5 for the two samples that will be discussed in this
subsection. Usually, the effect of a not-small-enough-voltage on the
measurement is in the same direction of raising the sample temperature. For
example, the resistance decreases with $V$ ( Fig.5), and so does the noise
magnitude (figure 7c), in both cases, the effect is monotonous with $V$. It
turns out that the degree of non-Gaussianity of the type considered here,
behaves in a qualitatively different way. In particular, it is non-monotonous
with $V,$ and it tends to vanish at both, high bias \textit{and} at low bias
thus peaking at an intermediate value of bias.

As a quantitative measure of the degree of non-Gaussianity we take the area
under the curve of the second spectrum as depicted in Fig.3, which we label as
$Int(S_{2})$. Using this scheme, we plot the dependence of this quantity on
$V$ for one of our smallest sample that exhibits non-Gaussianity (Fig.6). The
first set of data (solid circles) were taken initially without regard to the
order of changing $V~$just to test the effect produced$.$ When it was realized
that the correlations may disappear at high $V,$ the field was taken to a much
larger value ($V=46$ mV) to better define the asymptotic behavior of
$Int(S_{2})$. A later attempt to add more points to the curve failed to
reproduce the position of the peak in $Int(S_{2})$ versus $V$ obtained in the
first series. Rather, the curve seems to have shifted towards a lower bias.
The average sample resistance $r$ and its dependence on voltage $r(V)$ were
not affected by the high $V$ exposure. We shall return to this `mesoscopic'
effect following the discussion in the next section.

Figure 7 shows a more elaborate study of another $2x2\mu m$ sample with
similar noise characteristics except that now care was taken not to subject
the sample to an excessively large $V.$ The same pattern in terms of
$Int(S_{2})$ versus $V$ emerged (upper graph in Fig.7), and this time the
curve was fairly well reproduced by a second series of measurements. Along
with the second spectrum analysis, the figure shows the respective dependence
of the power-spectrum parameters $\eta$ and amplitude. Neither shows the
non-monotonic $V$ dependence exhibited by $Int(S_{2}).$ It is natural that
above a certain field, both $Int(S_{2})$ and the noise magnitude decrease with
$V$; a large field, like temperature, decreases the hopping-length and
$\mathcal{L},$ which in turn means larger number of fluctuators to average
over. The tale-telling result is that, below certain field, $Int(S_{2})$
diminishes when $V$ decreases in a way that suggests a much smaller effect as
$V\rightarrow0$. In other words, the noise appears to be correlated
\textit{because the measurement is not in the linear response regime}.

A plausible scenario that leads to correlations between individual fluctuators
is based on two ingredients; A) the current in the system is carried by a
percolation-network \cite{5,6,7}, and B) the (average) frequency $\omega_{i}$
of a fluctuator $i$ is, among other things, a function of the local voltage
$V_{i}.$ The latter is true for a generic two-level-state whether of `atomic'
or `electronic' nature; The frequency of the two-level-state usually depends
exponentially on the local voltage \cite{15}, so this is a sensitive source of
inter-modulation once the local $V_{i}$'s are re-distributed by the DCR
effect. Ingredient A is an inherent feature of variable-range-hopping systems,
and it is probably a common feature in other disordered conductors as well.
The system may then be viewed as a random-resistor-network where the active
fluctuators are part of it (or situated nearby such that can modulate a
resistance that is in the current path). When a fluctuator $j$ in the network,
changes its state, the local voltage on fluctuator $i$ will change too due to
the continuity of the current carrying network. If the resulting voltage
change $\delta V_{i}$ is not much smaller than $k_{B}T,$ the switch of $j$
will result is a change $\delta\omega_{i}$ of fluctuator $i$ frequency. Such
an effect is the basic building-block of a hierarchical correlation chain
where the slower fluctuator modulates a faster one. Naturally, this mechanism
for non-Gaussianity must vanish with the applied field as indeed is observed.
It should also be negligible when the system size is much bigger than the
range of the proposed inter-modulation effect. These considerations will be
now dealt with in a more formal way, and the results will be compared with the experiments.

\subsection{Theoretical considerations and comparison with experiments}

We consider a 2D hopping system and use the standard percolation scheme
\cite{16}. Focusing attention on a given critical resistor $j$ which is
affected by some fluctuator. For simplicity we assume that each resistor is
coupled only to a single fluctuator and will denote this fluctuator by the
same index $j$. Two different candidates for the role of the basic fluctuators
may be considered. The first is an `atomic' two level system as in amorphous
materials (see, e.g., Mott and Davis \cite{17}). The second one is an
aggregate composed of localized sites (not necessarily part of the percolation
cluster), and having two metastable configurations characterized by different
distribution of the electrons over the sites. The simplest object of such a
sort is a pair of hopping sites occupied by a single electron considered as a
source of 1/f noise in \cite{18}. Although quantitatively the effects of
atomic and electronic two level systems are expected to be different and
depend on different parameters of the material, in both cases the fluctuators
are expected to be sensitive to the local electric fields (for the atomic two
level systems the corresponding coupling is related to a presence of electric
dipolar moment of the atomic two level system). The model considerations given
below are applicable for either mechanism, although the microscopic equations
for the coupling coefficients are naturally different.

The fluctuation of the corresponding voltage swing, $\delta V_{j}$, will
inevitably lead to a fluctuation of the voltage across all other resistors.
This is the DCR effect considered by Seidler et al \cite{2} and shown to lead
to a colored second-spectrum of magnitude $S_{2,DCR}$. Next we show that in
the hopping system the voltage re-distribution in the CCN results in a more
elaborate coupling mechanism between different fluctuators. In particular,
this results in a dependence of the switching rate of the fluctuator $i$ on
the state of the fluctuator $j$. This coupling gives an additional,
non-linear, contribution to the frequency dependent second spectrum
($S_{2,corr}$). At large bias both $S_{2,DCR}$ and $S_{2,corr}$ are suppressed
with $V$ due to an increase of the number of effective fluctuators $N$. We
will show that $S_{2,corr}$ initially increases with $V$ and peaks at some
intermediate $V$ consistent with the experimentally observed behavior (figures
6 and 7).

Denoting by $\delta V_{ji}$ the voltage swing on resistor-$i$ affected by the
corresponding variation of resistor-$j,$ and assume that the main ensuing
effect is a variation of $\Delta_{i},$ the difference in energy between the
two states of the fluctuator;
\begin{equation}
\delta(\Delta_{i})=B_{i}\delta V_{ji} \label{deltai}%
\end{equation}
where $B_{i}$ is a coupling coefficient. In terms of the correlation length
$\mathcal{L}$, and for $R_{ij}\gg\mathcal{L}$ ($R_{ij}$ is the distance
between the critical resistors $i$ and $j)~\delta V_{ji}$ can be estimated
as:
\begin{equation}
\delta V_{ji}\sim V_{j}\frac{\delta G_{j}}{G_{j}}\frac{\mathcal{L}}{R_{ij}}
\label{Vij}%
\end{equation}
where in the regime $eV_{j}<<k_{B}T$ one has $|\delta G_{j}/G_{j}%
|=|\delta\varepsilon_{j}/k_{B}T|$, and $\delta\varepsilon_{j}$ is related to
the modulation of the resistor activation energy by the current re-distribution.

The variation of $\Delta_{j}$ leads to a variation of the fluctuator dwell
times. The latter can be written as \cite{19}
\begin{equation}
\tau^{-1}=\tau_{+}^{-1}+\tau_{-}^{-1}=\tau_{+}^{-1}\left(  1+\frac{n}%
{1-n}\right)  =\tau_{+}^{-1}(1-n)^{-1} \label{tau}%
\end{equation}
where $\tau_{+}$ and $\tau_{-}$ are the dwell times for the upper and lower
level, respectively, and n is the occupation at the upper-level state. Note
that $\tau_{+}$ corresponds to a transition accompanied by the emission of the
phonon with a frequency $\omega=\Delta/\hbar$. Thus, the change in $\tau_{+}$
due to the fluctuation of $\Delta$ is
\begin{equation}
\delta\tau_{+}\simeq-\tau_{+}\frac{\alpha}{\Delta}(\delta\Delta) \label{del}%
\end{equation}
where we have taken into account that $\tau_{+}^{-1}\propto\Delta^{\alpha}$.
Note that $\alpha\approx3$ for either a fluctuator of electronic nature in the
limit of small $\Delta_{i~}$\cite{19}, as well as for a fluctuator of atomic
nature \cite{20},

Combining eqs. \ref{tau},\ref{del} we obtain for the fluctuation of the
relaxation time of $i$-th fluctuator:
\begin{equation}
\delta\tau_{i}=-\tau_{i}\left(  \frac{\alpha}{\Delta_{i}}-\frac{n_{i}}{k_{B}%
T}\right)  \delta\Delta_{i} \label{deltatau}%
\end{equation}
Note that for an ideal $1/f$ first spectrum (which, strictly speaking, can be
realized only for $L\rightarrow\infty$), the fluctuations of $\tau_{i}$ would
not lead to a variation of $S_{1}$. However for a small sample size (where
deviations from $1/f$ spectrum may be observable as, e.g., Fig.1), the first
spectrum may be significantly affected by fluctuations of $\tau_{i}$. In the
limit $eV_{j}<<k_{B}T$ (where $\mathcal{L}=\mathcal{L}_{0}$) the effect is
proportional to $V_{j}$ and thus to the total bias $V$ ($V_{j}\approx
\frac{V\mathcal{L}}{L})$.

It can be shown (see Appendix 1) the resulting contribution to $S_{2}$ can be
estimated as
\begin{equation}
S_{2,corr}\propto\alpha^{2}{B}_{i}\delta V_{ij}\propto V \label{s2corr}%
\end{equation}

By comparison, the contribution of the DCR is given as
\begin{equation}
S_{2,DCR}\propto T\bar{\gamma}_{ij} \label{s2DCR}%
\end{equation}
Here
\[
\gamma_{ij}=\frac{\delta V_{ij}^{2}}{V_{i}^{2}}%
\]
which at $eV_{i}<<k_{B}T$ does not depend on $V$ while $\bar{\gamma}_{ij}$
means an average magnitude of $\gamma_{ij}$; the prefactors in eqs.
\ref{s2corr},\ref{s2DCR} differ from one another only by a numerical
coefficient of the order of unity.

Eqs.\ref{s2corr},\ref{s2DCR} then suggest that $S_{2,corr}>S_{2,DCR}$ if
\begin{equation}
e{\bar{V}_{i}}>k_{B}T\frac{\bar{\gamma}_{ij}}{\bar{B}_{i}\alpha^{2}}%
\end{equation}
In other words, our mechanism dominates over the DCR even when $e{\bar{V}_{i}%
}\ll k_{B}T$ provided
\begin{equation}
{\bar{\gamma}_{ij}}<\alpha^{2}{\bar{B}}_{i} \label{gama}%
\end{equation}
Recall that $B_{i}$ describes a relative effect of the fluctuation of the
resistor potential on the state of the nearby fluctuator while ${\bar{\gamma
}_{ij}}$ describes the relative fluctuation of the voltage on the resistor $i$
due to a fluctuator modulating resistor $j$. In general, $R_{ij}>\mathcal{L}$
and this ratio is small. At the same time the effect of the resistor voltage
on the fluctuator can be large enough. It holds in particular for the
fluctuators of the electronic origin if their size is comparable to the
inter-site distance within the hopping resistor $r_{h}$ provided that it is
situated at the distances less or comparable to $r_{h}$. It also holds for the
structural fluctuators provided they are situated close enough to the site
with a lower energy. In these cases $B_{i}\sim1$ therefore condition
\ref{gama} may be obeyed.

Now let us discuss the regime ${\bar{V}}_{i}\geq k_{B}T$ when the hopping
conductivity is strongly non-linear. It can be shown that the fluctuation of
the conductance of resistor $i$ resulting from the fluctuation of its
activation energy for $eV_{i}>k_{B}T$ is still given by $dG_{i}\simeq
-G_{i}(d\varepsilon_{i})/k_{B}T$. However the relative fluctuation of the
total conductance of the sample depends on the correlation length
$\mathcal{L}$:
\begin{equation}
S_{1}(\omega)\equiv\frac{(dG,dG)_{\omega}}{G^{2}}=\frac{\mathcal{L}^{4}}%
{L^{4}}\sum_{i}\frac{(dG_{i},dG_{i})}{G_{i}^{2}} \label{S1}%
\end{equation}
where $L$ is a linear size of the sample while $\mathcal{L}$ is the
correlation length of the percolation cluster which in the nonlinear regime
can be estimated as \cite{21}
\begin{equation}
\mathcal{L}\simeq\mathcal{L}_{0}\left(  \frac{k_{B}T}{eE\mathcal{L}_{0}%
}\right)  ^{\nu/(1+\nu)} \label{calL}%
\end{equation}
where $E$ is an average electric field within the sample, $\mathcal{L}%
_{0}=\mathcal{L}(V\rightarrow0),$ and $\nu$ is the percolation theory index
(for 2D $\nu\sim4/3$). Thus $\mathcal{L}\propto V^{-\nu/(1+\nu)}=V^{-4/7}$.
Correspondingly, one has
\begin{equation}
S_{1}(\omega)\propto\frac{\mathcal{L}^{2}}{L^{2}}\propto V^{-8/7} \label{S1V}%
\end{equation}
(we have taken into account that the result of a summation over the effective
resistors is proportional to the number of these resistors).

This ensemble-averaging effect is expected to be even stronger on the second
spectrum since it is a convolution of the two first spectra. So one expects
$S_{2}\propto N^{-2}$ where $N=(L/\mathcal{L})^{2}$ is the number of
fluctuators, then:
\begin{equation}
S_{2,corr}\propto\frac{V}{N^{2}(V)} \label{S2amp}%
\end{equation}
This is strictly obeyed for $eV<k_{B}T,$ at higher voltages there may be
contributions of other non-linear mechanism not considered here. $\ $Note that
Eq.13 contains two factors. The numerator ($V$) depends on the coupling
coefficient given in Eqs. 1-2 (see Appendix) and is indeed defined for the
case $eV<k_{B}T$. For larger bias this coupling coefficient is suppressed due
to the non-linearity of the medium and corresponds to a sub-linear behavior;
we do not consider this effect in detail. The denominator depends on the
statistical average discussed above Eq.13, and it remains the same for
$eV>k_{B}T$ while its dependence on $V$ follows from the considerations given
in Eq.11.

To compare, the respective contribution of the DCR mechanism has the following
dependence on $V:$
\begin{equation}
S_{2,DCR}\propto\frac{1}{N^{2}(V)} \label{S2DCRamp}%
\end{equation}

Let us now see how these expectations compare with our experiments. To find
the qualitative dependence on $V$ of the second-spectrum amplitude we need to
know the function $N^{2}(V),$ namely how the number of fluctuators varies with
$V$ over the range relevant for the experiment. This may be estimated
theoretically using similar considerations as those that led to eq.\ref{S1V}
above. The dependence of the first spectrum amplitude on $V$ is in rough
agreement with this equation (c.f., Fig.7c). However, the data for the second
spectrum amplitude were taken over more extensive range of $V,$ exceeding the
limits of validity of the power-law relation expected by eq.\ref{S1V}. It is
therefore necessary to get an estimate for $N^{2}(V)$ from experiments. That
can be done through the use of data for the conductance fluctuation versus
$V~$such as the results shown in Fig.8a$.$ To construct an empirical
$N^{2}(V)$, one then uses the relation $\frac{\Delta G}{G}(V)\approx
\sqrt{\frac{\mathcal{L}^{2}}{LW}}~=N(V)^{-\frac{1}{2}}$ discussed in section
2. This procedure yields the $N(V)$ depicted in Fig.8b, which empirically,
fits rather well an exponential dependence; $N(V)\propto\exp(\sqrt{V})$
\cite{22}. Using this form in eqs.\ref{S2amp},\ref{S2DCRamp} one gets the
qualitative dependence on $V$ for the two mechanisms for the second-spectrum
considered here. This is schematically illustrated in Fig.9. The overall shape
in this plot is in fair agreement with the experimental curves (c.f., Fig.6
and 7) although it seems that the small bias regime would fit better a faster
than linear with $V$ relation. To estimate the value of the voltage $V_{i}$
(that at the peak of $Int(S_{2})$ versus $V$ should probably be of the order
of $k_{B}T,$ c.f., Fig.9), one needs to know the number of critical resistors
in the sample. As noted above, this can be done based on the relative
magnitude of the CF, namely, the reproducible fluctuations in $G(V_{g})$
generated by sweeping the gate voltage $V_{g}$. The number of critical
resistors along the sample in Fig.6 can be be estimated using the data in
Fig.8b. Note that the applied voltage at the peak of $Int(S_{2})$ is
$V_{p}\approx5mV~$and$~V_{p}\approx10mV~$for the data in squares and circles
respectively (c.f., Fig.6)$.$ The value $N$ for $V\approx10mV$ can be read
from Fig.8b as $N\approx260,$ which then means that the value of $V_{i}$ at
the peak is $\approx\frac{10}{16}mV\approx0.6mV.$ Similarly, The value $N$ for
$V\approx5mV$ is $N\approx91,$ giving $V_{i}$ $\approx\frac{5}{9.6}%
mV\approx0.52mV.$ These values compare favorably with the sample temperature
$T=4.11K$.

We were not able to measure noise in the 2$\mu m$ samples using bias that is
strictly in the linear response regime ($eV_{i}\ll k_{B}T$). In fact,
deviations from linear response in the range of bias used here are reflected
in the sample conductance itself (Fig.5). The smallest bias used for the
sample in Fig.6 was $V=1.4mV$ which, in terms of $V_{i}$, is the equivalent of
$3K$ (using the respective $N\simeq29$ $~$from Fig.8. This bias is not much
smaller than the bath temperature. Nevertheless, the low bias regime we did
manage to use is low enough to expose the peak in $Int(S_{2})$ vs. $V$
consistent with the proposed mechanism.

Note that the relevance of both correlated-noise scenarios discussed here
hinges on specific assumptions. The DCR assumes that the conductance swings
associated with slow fluctuators are potent enough to give a significant
contribution. The non-linear mechanism we offered requires the existence of
`soft' fluctuators that, in addition, are coupled effectively to critical
resistors. In either case the `master fluctuator(s)' should operate on the
frequency window that is relevant for the experimental scales. In the regime
of mesoscopic samples one may expect to find fluctuators realizations such
that some (or all) of these conditions are not realized in which case the
non-Gaussianity will be weak or absent. Moreover, applying a voltage may
displace a key fluctuator out of its `commanding' position thereby removing
the origin of correlations. This is certainly a concern in the strongly
non-linear regime. Applying a large field will inevitably modify the
current-carrying network. The modification may be reversible, in which case we
expect that data such as in figures 6 and 7 will reproduce themselves under
$V$ cycling. However, when the applied $V$ is sufficiently large, it is quite
likely that a different percolation network will be precipitated, just as a
thermal cycling involving high temperatures would cause \cite{23}. This may
lead to noise data of a different statistical nature. The change may well be
subtle; turning on or off the coupling of certain fluctuators to the CCN is
all that is needed. The average disorder may not change in the $V$ - recycling
process, and the CF pattern may be only slightly affected. We believe that
these considerations account qualitatively for the non-systematic occurrences
of non-Gaussian noise in our samples, as well as for the effect of $V$ cycling
discussed in section II.

It should be mentioned that there are other non-linear mechanisms, not
considered here, that may contribute to correlations between remote
fluctuators, especially at the high $V$ regime. For example, the change of the
local voltages could modify the values of the critical resistors in the CCN
\cite{24}, and may give rise to new fluctuators. Another interesting scenario
is a heterodyne effect; namely, frequency mixing of the `master' frequency
with the `local' one due to the non-linearity of the critical resistors. All
such mechanisms, as well as the DCR, should be seriously considered whenever a
noise with colored second-spectrum is encountered in a conducting system. Note
that a generic feature of these mechanisms is their long range $1/R$ nature
(see eq.\ref{Vij} ) making them more effective than most `direct'
interactions. As was demonstrated in this work, these effects may bring about
correlations between fluctuators even in samples that are considerably larger
than the scale relevant for the interactions associated with the
electron-glass \cite{23}.

Finally, it should be mentioned that while reducing the applied $V$ below the
value where $Int(S_{2})$ peaks diminishes the non-Gaussian effect, the glassy
effects if anything, become more prominent \cite{23,25}. Clearly then,
\textit{glassiness and correlated-noise (when exists) are not necessarily
related}.

In summary, we have described a set of conductance-noise experiments on
disordered films of $In_{2}O_{3-x}$ in their glassy phase. The emphasis in
this study was on the degree of noise correlation as function of system size.
Noise correlation was measured by the second spectrum of conductance data
taken at liquid helium temperatures. Our main finding is that, down to sample
size of $2\mu m$ the noise has the usual $1/f$ power-spectrum. Hopping samples
with this size contain small number of critical resistors as indicated by the
prominent conductance-fluctuations they exhibit. Such samples still show
essentially all the electron-glass features as macroscopic samples \cite{23}.
Given the way non-Gaussianity decreases below a certain bias in these samples
it seems unlikely that the correlations observed at finite bias are due to glassiness.

When the voltage used in the conductance measurements was not small enough,
non-Gaussian noise was observed in several of the samples, including samples
as large as $40\mu m.$ We have demonstrated that the degree of non-Gaussianity
is a non-trivial function of the bias. A model that purports to account for
these findings was offered, and its consequences are found to be in
qualitative agreement with our experiments.

A lesson that may be taken from our study is that correlations in conductance
noise may arise from a non-linear mechanism, and this may be of particular
relevance to disordered conductors measured at low temperature. Such effects
need be better understood and carefully examined before a non-Gaussian noise
is associated with correlations due to, e.g., glass. As a point of principle,
one indeed expects that, on sufficiently small scales, a glassy system may
show a correlated noise. What should perhaps be stressed is that the converse
is not necessarily true.

We acknowledge illuminating discussion with Clare Yu on her computer
simulation results of the DCR model. One of us (V.I.K.) acknowledges support
of the Lady Davis Foundation. This research has been supported by the
Binational US-Israel Science Foundation and by The Israeli Academy for
Sciences and Humanities.

\section{Appendix}

Here we consider the coupling between the fluctuators in more detail, and
estimate the effect of this coupling on the second spectrum. For simplicity we
assume that the (initially un-correlated) `fast' fluctuators are coupled to
`slow' ones, and the coupling is characterized by the occupation numbers
$\tilde{n}_{j}$. By 'fast' and 'slow' fluctuators we refer to specific
two-level-systems that contribute in the measured noise, (at the high end of
the fluctuators vs. frequency distribution and at the lower end respectively).
Explicitly we consider only the modulation effect of 'slow' fluctuators on
'fast' one. The complementary process (i.e., 'fast' affecting 'slow') while
possible is much more involved and is not treated here.

The `slow' fluctuators affect the value of $\Delta_{i}$ of the `fast'
fluctuators. A natural result of this modification is change of the
fluctuators occupation numbers, $n_{i}$.

Making use of Eq.\ref{deltai}, and using $n=(\exp(\Delta/k_{B}T)+1)^{-1}$ one
concludes that in a presence of `slow' fluctuator $j$ one has
\begin{equation}
n_{i}(1-n_{i})|_{t}=n_{i}(1-n_{i})|_{0}(1-\frac{\tanh(\Delta_{i}/2k_{B}%
T)}{2k_{B}T}{\tilde{n}}_{j}(t)\delta\Delta_{i,j})
\end{equation}

Another source of the fluctuations is related to fluctuations of the
relaxation time $\tau_{i}$. One readily obtains:
\begin{equation}
\delta\left(  \frac{\tau_{i}}{1+(\omega\tau_{i})^{2}}\right)  =\delta(\tau
_{i})\frac{1-(\omega\tau_{i})^{2}}{(1+(\omega\tau_{i})^{2})^{2}}%
\end{equation}
Thus, the contribution of $i-th$ resistor to the first spectrum is:
\begin{gather}
\delta\left(  (\delta n_{i},\delta n_{i})_{\omega}\right)  =A_{ij}{\tilde{n}%
}_{j};~\label{dela}\\
A_{ij}=-(\delta n_{i},\delta n_{i})_{\omega}|_{0}\frac{\tanh(\Delta_{i}%
/2k_{B}T)}{2k_{B}T}{\tilde{n}}_{j}(t)B_{i}\delta V_{i,j}-\nonumber\\
-\tau_{i}\frac{1-(\omega\tau_{i})^{2}}{(1+(\omega\tau_{i})^{2})^{2}}%
n_{i}(1-n_{i})\left(  \frac{\alpha}{\Delta_{i}}-\frac{n_{i}}{k_{B}T}\right)
B_{i}\delta V_{ij}\nonumber
\end{gather}
As a result, the first spectrum appears to be dependent on the occupation
number of the `slow' fluctuators.

Now let us estimate the effect of the correlations on the second spectrum
$S_{2}(\omega_{2})$. By definition for $S_{2}$ we have
\begin{gather}
S_{2}(\omega_{1},\omega_{2})=G^{-4}\int_{0}^{t_{max}}d\tau e^{i\omega\tau}%
\int_{0}^{t_{max}}dt\nonumber\\
\cdot\int_{t}^{t+t_{0}}dt^{\prime}\int_{t^{\prime}}^{t^{\prime}+t_{0}}%
d\tau^{\prime}\delta G(t^{\prime}+\tau^{\prime})\delta G(t^{\prime}%
)e^{i\tau^{\prime}\omega_{1}}\nonumber\\
\cdot\int_{t+\tau}^{t+\tau+t_{0}}dt^{\prime\prime}\int_{t^{\prime\prime}%
}^{t^{\prime\prime}+t_{0}}d\tau^{\prime\prime}\delta G(t^{\prime\prime}%
+\tau^{\prime\prime})\delta G(t^{\prime\prime})e^{i\tau^{\prime\prime}%
\omega_{1}}%
\end{gather}
where in our case
\begin{equation}
\delta G=\sum_{i}g_{i}\delta n_{i} \label{g}%
\end{equation}
Thus, the spectrum of the fluctuations is related to the temporal behavior of
$\delta n_{i}$. Here $g_{i}$ are the coefficients describing a contribution of
$i$-th fluctuator to the conductance fluctuations. Let us first consider a
case of statistically independent fluctuators. In this case
\begin{equation}
\delta G(t^{\prime})\delta G(t^{\prime}+\tau^{\prime})=\sum_{i}g_{i}%
^{2}(\delta n_{i}(t^{\prime})\delta n(t^{\prime}+\tau^{\prime}))
\end{equation}

The DCR mechanism results from the modulation of the coefficient $g_{i}$ in
Eq.\ref{g} by another (`slow') fluctuator $j$. The corresponding contribution
to $S_{2}$ can be estimated as
\begin{gather}
S_{2,DCR}\propto\sum_{i,j}\gamma_{ij}(\delta n_{i},\delta n_{i})_{\omega_{1}%
}^{2}(\delta{\tilde{n}}j,\delta{\tilde{n}}_{j})_{\omega_{2}}=\nonumber\\
{\bar{P}}k_{B}Tf_{2}(\omega_{1}){\bar{\gamma}_{ij}}\sum_{j}(\delta{\tilde{n}%
}j,\delta{\tilde{n}}_{j})_{\omega_{2}}%
\end{gather}
In contrast, in our case we deal with real coupling between the fluctuators
described by Eq.\ref{deltai} which exists only at finite $V$. Using
Eq.\ref{dela} one obtains a contribution to $S_{2}$ $\propto\sum_{ij}%
A_{ij}^{2}(\delta{\tilde{n}}_{j},\delta{\tilde{n}}_{j})$. The most important
term here is the one $\propto(\alpha/\Delta_{i})^{2}$. Indeed, due to a
presence of `soft' fluctuators with $\Delta_{i}\rightarrow0$ this term is
divergent. The divergency has a clear cut-off at $\Delta_{i}\sim\delta
\Delta_{i}=B_{i}V_{ij}$ since in our calculations we assumed the fluctuations
of $\Delta_{i}$ to be small (note that here we assume $eV_{i}<k_{B}T$).
Summing over the different fluctuators gives
\begin{equation}
\sum_{i}\frac{(\delta\Delta_{i})^{2}}{(\Delta_{i})^{2}}\simeq(B\delta
V)^{2}\int_{\delta\Delta}d(\Delta)P(\Delta)(\Delta)^{-2}={\bar{P}}B\delta V
\label{si}%
\end{equation}
where we have assumed that the distribution $P(\Delta)=(\bar{P})$ is flat.
Thus here the summation over different $\Delta$ is controlled by the lower
limit of `soft' fluctuators, and we obtain
\begin{equation}
S_{2,corr}\propto\alpha^{2}{\bar{P}}{\bar{B}_{i}}{\bar{\delta}V_{ij}}%
f_{1}(\omega_{1})\sum_{j}(\delta{\tilde{n}}j,\delta{\tilde{n}}_{j}%
)_{\omega_{2}} \label{s2corr1}%
\end{equation}

Assuming that the `slow' fluctuators have exponentially broad scatter of
relaxation times, the contribution to the second spectrum is $S_{2,corr}%
\propto\omega_{2}^{-1}$ (at the frequency scale $t_{0}^{-1}>\omega_{2}%
>t_{max}^{-1}).$

Note that the non-linear effect we propose will dominate over the DCR
mechanism even for moderate bias $eV_{i}<k_{B}T$. The reason is that among
different fluctuators there exist those with small $\Delta_{i}$. For such
fluctuators the relative modulation of the relaxation rate due to fluctuations
of $\Delta_{i}$ as it is seen from Eq. \ref{deltatau} appears to be
$\propto\delta\Delta_{i}/\Delta_{i}$. Since the corresponding contribution to
$S_{2}$ is proportional to $(\delta\Delta_{i}/\Delta_{i})^{2}$ it leads to a
significant enhancement of a role of `soft' fluctuators with small
$\Delta_{i~}$as is indicated by eq. \ref{si}.

\section{Figures captions}

FIG. 1. Conductance noise power-spectrum measured under bias voltage $V=3$mV.
The lower frequency part of the spectrum represents the averaging over 16
Fourier transformed $G(t)$ runs. The higher frequency part of the
power-spectrum was averaged over 1024 power-spectrum curves measured on the
resistor $R=1.01M\Omega$, connected in series with the sample, employing
HP35670A. The dashed line depicts $1/f^{1.08}$ dependence (a best fit to the
data). Inset: typical $G(t)$ series used for calculation of the lower
frequency part of the power-spectrum. Note the relatively low signal to noise
ratio typical for mesoscopic samples. The sample: length=$2\mu m$, width=$2\mu
m$, $R_{\square}=5M\Omega$.

FIG. 2. The magnitude of the noise power-spectrum measured at $f=10mHz$ as a
function of sample area $A$ ($=LxW$). Each datum point represents the
averaging over several samples with same area and values of the resistance
ranging between $R_{\square}=2M\Omega$ and $R_{\square}=10M\Omega$. The dashed
line depicts the $S_{1}\propto A^{-1}$ law.\newline

FIG. 3. Second-spectrum as a function of normalized frequency measured in two
samples with the same lateral dimensions; $L=30\mu m$, and $W=40\mu m$, and
resistances; $R_{\square}=4.8M\Omega,$ and $R_{\square}=5.3M\Omega$ for
samples in \textbf{(a)} and \textbf{(b)} respectively. The octaves (2 to 9)
are labelled by the corresponding lowest frequency values ($f_{L}$) and
represented by different symbols on the plot. The measurements for both
samples were performed using a series resistor $R=100k\Omega$ and applying
$V=100$mV, and $V=150$mV as a bias voltage for the sample in \textbf{(a)} and
\textbf{(b) }respectively. The degree of non-Gaussianity (labeled in this work
as $Int(S_{2}),~$see text) is taken to be proportional to the area defined by
the dashed and dotted lines, and the ordinate axis (plate \textbf{a}).

FIG. 4. The `memory' cusp (see reference 9) as seen in the measurements of
conductance as a function of the gate voltage for the sample shown in
Fig.3\textbf{(b)}. Two successively measured traces (solid and open circles)
show reproducible conductance fluctuation (CF). Gate voltage scan rate was
$0.02V/sec$, bias voltage $V=20$mV.\newline

FIG. 5. Conductance as a function of the applied bias measured for two samples
with the same lateral dimensions; $L=2\mu m$, $W=2\mu m$. Note the deviation
from linear response even at the smallest bias used.\newline

FIG. 6. The degree of non-Gaussianity $Int(S_{2})$ as function of the applied
bias for two series of measurements: prior to (solid circles), and after (open
squares) the application of high bias (see text). Dashed lines are guides for
the eye. The measurements were performed using a series resistor
$R=1.01M\Omega$. The sample: $L=2\mu m$, $W=2\mu m$, $R_{\square}=5M\Omega$.

FIG. 7. Plate \textbf{a}: The degree of non-Gaussianity as function of the
applied bias in two series of measurements: successively increasing the bias
(open circles), and a later set, employing bias values within the same range
as before with no particular order (solid circles). The value of $\eta$ (in
the first-spectrum law, $1/f^{\eta}$), and the noise power per decade as
function of the applied bias - \textbf{(b)} and \textbf{(c)}, respectively.
Dashed lines are guides for the eye. The measurements were performed using a
series resistor $R=1.01M\Omega$. Sample parameters: $L=2\mu m$, $W=2\mu m$,
$R_{\square}=9M\Omega$.\newline

FIG. 8. Plate\textbf{ (a): }Conductance versus gate-voltage scans taken with
different values of bias $V~$for the sample in Fig.6$.$ Each trace shows
reproducible conductance fluctuations with rms amplitude that decreases with
$V.$ The average value of the conductance increases with bias (c.f., Fig.5).
Plate\textbf{ (b):} The values of $N$ for the series of traces shown in
\textbf{(a)} estimated by the rms amplitude of the fluctuations (see text).

FIG. 9. A schematic dependence of $S_{2,corr}$ and $S_{2,DCR}$ on the applied
voltage (see text).


\begin{thebibliography}{99}
\bibitem {22}Note incidentally that, for comparable applied fields ($=V/L$),
the CF in the sample in Fig.8 are much larger than the fluctuations of the
$30\mu m~$by $40\mu m$ sample of Fig.4, consistent with the $\frac{\Delta
G}{G}\approx\sqrt{\frac{\mathcal{L}^{2}}{LW}}$ relation.
\bibitem {24}This non-linear effect was not considered in reference \cite{2}.
\end{thebibliography}
\end{document}